\documentclass[fleqn,10pt]{olplainarticle}
\usepackage{caption, csquotes, epigraph, hyperref}
\title{Aphorisms on Epidemiological Modelling}

\author[1]{Juan Afanador}
\affil[1]{jafanad@gmail.com}


\begin{abstract}
Epidemiological modelling is critiqued towards a scientific practice of negativity in the context of Scotland’s Centre of Expertise on Animal Disease Outbreaks (EPIC). The paratactical approach to the melancholy science is invoked to problematise One Health, the intra-pandemic modelling culture, and to delineate an inkling of the negative in EPIC's work. 
\end{abstract}

\begin{document}

\flushbottom
\maketitle
\thispagestyle{empty}

\epigraph{\textit{Come closer}}{\cite{adorno1993messages}}

\section{``The transcendence of the One evokes the aspiration to a wisdom that is not knowledge, [\ldots] that is Love" \citep{levinas1999alterity}}
One Health is a collection of learnings and practices geared towards the reproduction of our social specificity and the administration of life (v. \cite{zinsstag2020one}). Expressed in the actor-network idiom (v. \cite{boden2020epic}), ``\textit{One Health [\ldots] address[es] a health threat at the human-animal-environment interface based on collaboration, communication, and coordination across all relevant sectors and disciplines, with the ultimate goal of achieving optimal health outcomes for both people and animals}'' \citep{world2019taking}. The key operator of this double movement traversing ontological (humans/animals/environment) and societal (productive sectors and disciplines consecrating the bourgeois division of labour) domains is the undeducible notion of integrative oneness resting on an imperative of efficiency and progress. 

The One Health framework is the recuperation of a classicism. The rationality of One Health knowledge corresponds to the absoluteness of unity (the One) as praxis incarnate. One Health refounds the neo-Platonic One onto the unity of the system and the immanence of transcendental unity. \textit{Vide}, e.g., the merger of quantitative microbial risk assessment and material flow analysis for wastewater-related disease risks \citep{nguyen2009improving}, the ``farm to fork'' pathogen spread models \citep{nauta2007risk}, or proposals of integrated risk assessment \citep{racloz20158} -- they all harbour the promise of technical prolongations of great productive/consumptive potency.

\textit{Pace} \cite{zinsstag2020one},\cite{STEELE2021100314},\cite{streichert2022participation}, and \cite{boden202163}, this does not mean that One Health constitutes a theoretical development -- it is a cover-concept for an amalgamation of (ideologically determined) practices re-tooled around the historical contributions of veterinary and human medicine to the valorisation of value:

\begin{displayquote}
``\textit{One Health can thus be defined as any added value in terms of health of humans and animals, financial savings or environmental services achievable by the cooperation of human and veterinary medicine when compared to the two medicines working separately}'' \citep{zinsstag2020one}
\end{displayquote}

But maybe there is no need for a theoretical or conceptual grounding of One Health. Maybe the factive merits of human and veterinary medicine require no conceptual probing. Maybe we should only problematise the formal shortcomings of the framework -- i.e., asking whether we should shift from population-level (ODE-based) models to agent-based models (e.g. from \citep{anderson1992infectious} to \citep{roche2011agent}), or decisively concerning ourselves with the framework's ethical and moral reverberations \citep{boden2020epidemiology}. This, however, would obfuscate our ability to think the reality of the capitalist power inherent in scientific praxis, our ability to think the problematic genealogy of One Health and its \textit{plenum} of congenial frameworks (Planetary/Eco Health/One Medicine?), and our capacity to challenge the naturelikeness primacy of value blocking a categorical critique of gender and race.

Not that every One Health problematique can be collapsed into the value category -- notably the manifestations of the ancestral gender-based dissociation of sociality. We could ask whether the framework's totalising \textit{ethos} (or contribution thereof via the valorisation of value), would not leave room for the particularity of beings it covers? If, say, a  biological totality shows what makes up the concrete reality of a singular being -- thus, signifying that reality is heterogeneous -- could not perhaps the oneness of One Health illuminate particularities? If so, the framework would have moved beyond itself, replacing systems with constellations, and prefiguring a contestation of its immanent positivities. 

\section{``The I has its unfreedom demonstrated to it, within itself, by something alien to it -- by the feeling that ‘this isn't me at all’" \citep{adorno2003negative}}

Research programmes are devised to comply with a seemingly objective function -- that of unveiling the inner workings of reality, and providing the means to administer life (v. \cite{woods2017animals} and \cite{maureira2018epidemiological}). Health sciences occur in a historical specificity which determines their emergence and evolution. Health sciences also shape these conditions retroactively, through the affirmation of instrumental rationality and the forms of social praxis with which rationality is coeval. 

Scientific activity revolving around an abstract notion of health -- necessarily/implicitly -- effects the depersonalisation of wellbeing in its aspiration to universal analgesia. As such, Epidemiology studies the determinants and patterns of the wrong state of things, in accordance with the methods and techniques of an empiricism that believes to have arrived at unmediated -- i.e. true -- facts. Not a literal-minded empiricism, though, Epidemiology couches this wrong state of things in terms of dynamic systems, networks or combinations of the two \citep{edoh2018network}. The rationale is that, since populations of abstract individuals are interwoven in layers of spatio-temporal and systemic interrelatedness, alterations in normality can be traced as deviations from stable states or fixed points \citep{krieger2005embodiment, enright2018epidemics}. 

Systems are problematic insofar as their claim to unity forces a \textit{de facto} instrumentalisation of their constitutive parts, but even more so as epidemiologists operate a recourse to a teleological explanation of their functioning, which endows systems with \textit{ad hoc} validating qualities \citep{adorno2003negative}. When formalised – i.e., when temporal considerations are reduced to sheaves of ODEs – dynamic systems become consecrated tools for the mining of truths that are later manufactured into policy guidelines. For these moments occur as unreflective instances of social praxis, the, e.g., many variations on the SIR model end up furthering scientistic ideology by relegating the scrutiny of material/economic realities to the subjective domain. When the role of these realities, in determining the relevance and worth of research themes and results, is obscured or externalised, any capacity to question policy-making, at the level of their underpinning ideological mechanisms, is effectively sublated. 

Similar observations carry over to data-driven and purely statistical models of epidemiological predictions. The constraints that gauge critical reflection are more stringent but less conspicuous. Truth mining is taken to an unmediated limit where computational heuristics are applied to the numerical resolution of analytical problems. Allegedly, the bridging of reality and policy is more readily achieved through these types of epidemiological models, so the moment of falsity here does not lie in the particular contents of the models, but in the positive (unreflective) manipulation of such contents. In this act of legerdemain, normative actions become fully dissociated from societal determination – an ideological denial of the inherent connection between knowledge and social conflicts, which passes unnoticed under layers of neural networks.

\section{\textit{Auf einem Beine stehen}}
Identity thinking is the epistemic error that propagates instrumental rationality -- its definitive and constitutive essence. Let us lapse into the instrumentality of consequential thinking to explore the analytical horizons of critique within a scientific domain amenable to EPIC's work. This is the mobilising element behind the creation of a  \textit{Sentiment-based Mathematical Model Repository for disease emergency decision-making, and long-term resilience to future One Health threats} (SER).

SER is an invitation ``to think unguarded thoughts'' \citep{adorno1999education}. SER is the computational articulation of these thoughts, and the thinking of their thinking, with respect to the \textit{status quo} subtending the EPIC register. SER intended to say something meaningful about the early COVID modelling culture in connection with public health policy, while constructing a relational account (graph-based repository) of the incumbent works in the process. 

From a linguistics perspective, SER's entry point into critique is pragmatics \citep{pagin2014critical, bender2019linguistic} -- SER sets out to understand how context affects meaning, by identifying implicit information in the creation of meaning within our current social specificity, i.e. the capitalist modernity in the age of COVID. To an extent -- and here lies one of SER's more prominent nonidentitarian moments -- this could be seen as the tackling of a series of NLP tasks involving the labelling and classification of annotated data, requiring the creation of custom layers and tagsets capable of capturing the meta-analytical -- i.e., critical -- angle behind our proposal. 

SER delineates the discursive aspects of the narratives underpinning distinct modelling approaches, along with their societal requirements/imperatives, by looking into the context in which the textual contribution is embedded, and by subjecting the identified discursive elements to critical inquiry. Put differently, SER not only aims at naming epistemic entities (errors) and propagating classificatory thinking, but at operating on them in order to reveal the internal contradictions (social aporias) whereof the entities are symptomatic or exemplars. 

The disclosure of these contradictions involves the formal analysis of the COVID corpus through the lens of a multi-valued logic -- in particular a non-classical logic that allows for valid contradictions and operates on a non-classical sense of negation (negativity). I argue that extending a \textit{Bidirectional Encoder Representations from Transformers} model (BERT) \citep{devlin2018bert} with neural convolutional networks (CNN) \citep{malmqvist2020determining}, representing argumentative interactions among distinct research groups/modelling narratives (in the spirit of, e.g., DAX \citep{albini2020dax} and DEAr \citep{cocarascu2020data}), provides an amenable environment to implement a non-classical proof theory capable of checking the validity of the identified inferences/contradictions. 

Negative instrospection is the crux of SER. Negative introspection is not algorithmic -- in logics, negativity is typically identified with the various axioms governing the application of a (classical) negation operator $\sim$. The possibilities of the negative are subsumed either to the connexivist paradigm or to the principle of explosion. Connexivist logics approach negation as cancellation, i.e., $\sim\phi$ deletes $\phi$, as to void $\phi\land\sim\phi$ of content \citep{priest1999negation}. The modern approach to negation distinctly treats negation as an excluding and exhaustive representation of what is believed to lay outside the negated concept, i.e., $\sim\phi$ is the complement of $\phi$, so that $\phi\land\sim\phi$ entails any arbitrary (and irrelevant) proposition within the corresponding universe of discourse. Note that the cancellation view can be recuperated from the latter explosion model through a semantic interpretation within a possible worlds framework \citep{priest2007paraconsistency}.

Some consider the modern approach to be the default mode of negation, however neither is equipped to deal with real abstractions, i.e., neither is equipped to operate on inconsistencies, contradictions, and the ensuing paradoxes. An alternative view on negation is granted by replacing the idea of absolute identity, essential to both the connexivist and the explosion models, with a notion of identification where the negative merely outlines a concept without full determination \citep{priest2007paraconsistency}. In this context, the contradiction $\phi\land\sim\phi$ has the same inferential status as any other proposition, i.e., it may, or not, be valid at times. This is key to thinking of thinking -- the point of articulation of formal critique; a relevant negation which is historically situated.  

Of particular relevance for the elucidation of a practice of negativity in a computational idiom, are the works of \cite{grivas2020not} on \textbf{negation detection}, of \cite{jimenez2020corpora} on \textbf{corpora annotated with negation}, and of \cite{raji2021ai} on \textbf{construct validity}.

One more point: if SER fails to evoke negativity, it would risk the reification of critical thinking.

\section{Critical Models}

\begin{displayquote}
``\textit{[Genuine] thinking begins as soon as it ceases to content itself with cognitions that are predictable and from which nothing more emerges than what had been placed there beforehand. The humane significance of computers would be to unburden the thinking of living beings to the extent that thought would gain the freedom to attain a
knowledge that is not already implicit} \citep{adorno2005critical}''. 
\end{displayquote}

I argue that SER could be extended to work as an analytical pipeline, using Broadwick (or an updated version thereof) \citep{o2016broadwick} as its backbone, and a Generative Pre-trained Transformer (GPT) \citep{radford2018improving} model as the cache of the interactions occurring at the meta-analytical level. 

\begin{figure}[h]
    \centering
    \includegraphics[width=\textwidth]{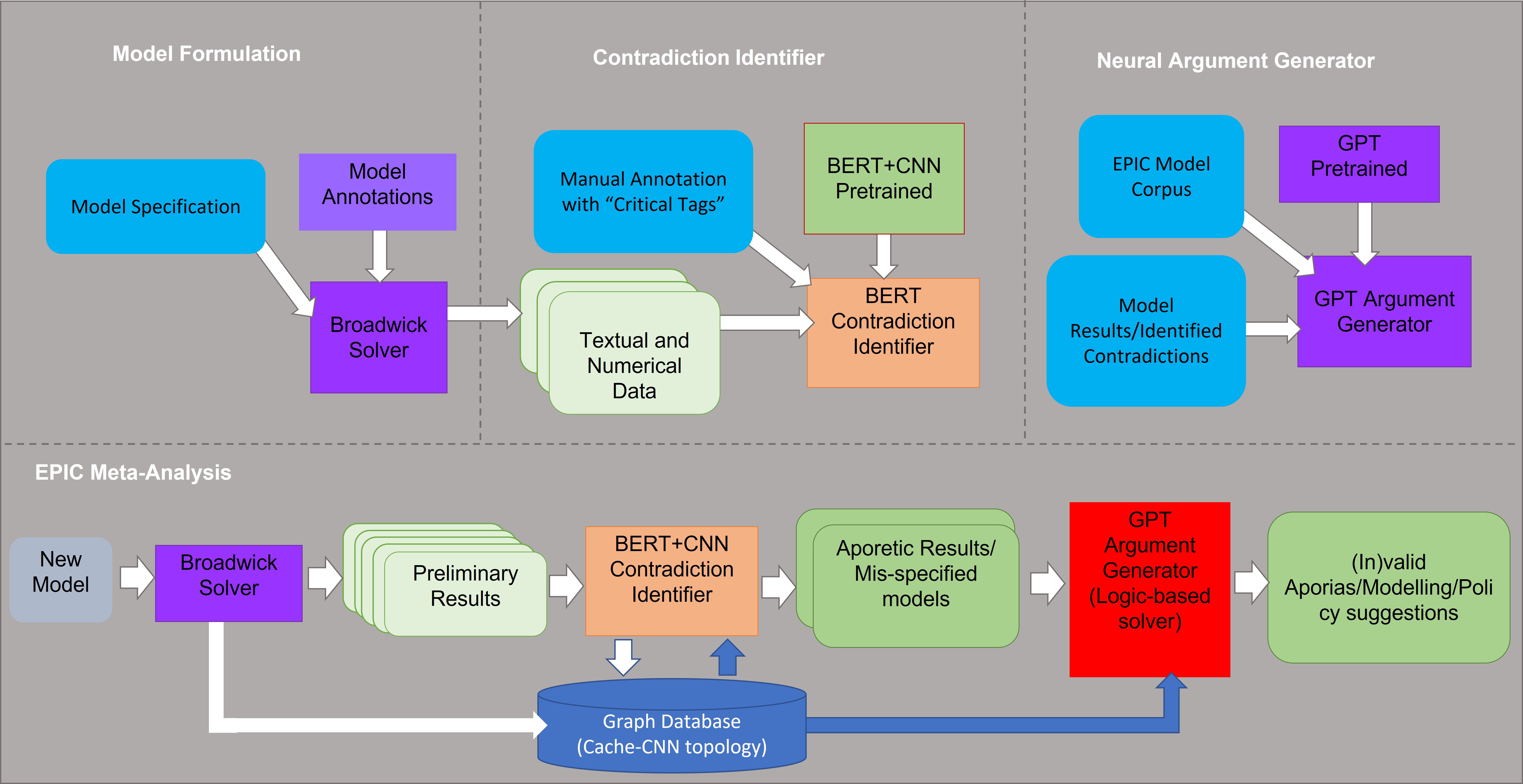}
    \caption{A Tentative EPIC Pipeline (loosely based on \cite{stedden2022})}
    \label{fig:epic_pipeline}
\end{figure}

Epidemiological models produced via Broadwick would be annotated through a multi-stage implementation of SER, involving a certain degree of manual annotation, the training of the BERT+CNN architecture which would perform any pending pragmatics tasks (i.e., the identification of contradictions), and the fine-tuning of a pre-trained argument generation model \citep{al2021employing}. This would provide the basis to critically assess the models' assumptions, reproduction/propagation of social aporias, and potential mis-specification/overfitting problems (See Figure \ref{fig:epic_pipeline}). The SER proof-of-concept operated on a Neo4j graph database \citep{needham2019graph}. This can still be the scaffolding around which to conceive an analytical pipeline. 

Many things are left unsaid. For SER to actualise the promise of what would be different, we need a theoretical development of \textit{cupiditas} and \textit{amor} (v. \cite{negri2013spinoza}) in the negative -- a computational articulation of our resistance to the social totality. 

\begin{thebibliography}{}

\bibitem[Adorno, 1993]{adorno1993messages}
Adorno, T. (1993).
\newblock Messages in a bottle.
\newblock {\em New Left Review}, 200:5--14.

\bibitem[Adorno, 2003]{adorno2003negative}
Adorno, T. (2003).
\newblock {\em Negative dialectics}.
\newblock Routledge.

\bibitem[Adorno, 2005]{adorno2005critical}
Adorno, T.~W. (2005).
\newblock {\em Critical models: Interventions and catchwords}.
\newblock Columbia University Press.

\bibitem[Adorno and Becker, 1999]{adorno1999education}
Adorno, T.~W. and Becker, H. (1999).
\newblock Education for maturity and responsibility.
\newblock {\em History of the Human Sciences}, 12(3):21--34.

\bibitem[Al~Khatib et~al., 2021]{al2021employing}
Al~Khatib, K., Trautner, L., Wachsmuth, H., Hou, Y., and Stein, B. (2021).
\newblock Employing argumentation knowledge graphs for neural argument
  generation.
\newblock In {\em Proceedings of the 59th Annual Meeting of the Association for
  Computational Linguistics and the 11th International Joint Conference on
  Natural Language Processing (Volume 1: Long Papers)}, pages 4744--4754.

\bibitem[Albini et~al., 2020]{albini2020dax}
Albini, E., Lertvittayakumjorn, P., Rago, A., and Toni, F. (2020).
\newblock Dax: Deep argumentative explanation for neural networks.
\newblock {\em arXiv preprint arXiv:2012.05766}.

\bibitem[Anderson and May, 1992]{anderson1992infectious}
Anderson, R.~M. and May, R.~M. (1992).
\newblock {\em Infectious diseases of humans: dynamics and control}.
\newblock Oxford university press.

\bibitem[Bender and Lascarides, 2019]{bender2019linguistic}
Bender, E.~M. and Lascarides, A. (2019).
\newblock Linguistic fundamentals for natural language processing ii: 100
  essentials from semantics and pragmatics.
\newblock {\em Synthesis Lectures on Human Language Technologies},
  12(3):1--268.

\bibitem[Boden, 2021]{boden202163}
Boden, L. (2021).
\newblock 63. modelling for pandemic preparedness: A need for a one health
  approach.
\newblock {\em Animal-science proceedings}, 12(1):48.

\bibitem[Boden and Mellor, 2020]{boden2020epidemiology}
Boden, L. and Mellor, D. (2020).
\newblock Epidemiology and ethics of antimicrobial resistance in animals.
\newblock In {\em Ethics and Drug Resistance: Collective Responsibility for
  Global Public Health}, pages 109--121. Springer, Cham.

\bibitem[Boden et~al., 2020]{boden2020epic}
Boden, L.~A., Voas, S., Mellor, D., and Auty, H. (2020).
\newblock Epic, scottish government's centre of expertise in animal disease
  outbreaks: A model for provision of risk-based evidence to policy.
\newblock {\em Frontiers in veterinary science}, 7:119.

\bibitem[Cocarascu et~al., 2020]{cocarascu2020data}
Cocarascu, O., Stylianou, A., {\v{C}}yras, K., and Toni, F. (2020).
\newblock Data-empowered argumentation for dialectically explainable
  predictions.
\newblock In {\em ECAI 2020}, pages 2449--2456. IOS Press.

\bibitem[Devlin et~al., 2018]{devlin2018bert}
Devlin, J., Chang, M.-W., Lee, K., and Toutanova, K. (2018).
\newblock Bert: Pre-training of deep bidirectional transformers for language
  understanding.
\newblock {\em arXiv preprint arXiv:1810.04805}.

\bibitem[Edoh and MacCarthy, 2018]{edoh2018network}
Edoh, K. and MacCarthy, E. (2018).
\newblock Network and equation-based models in epidemiology.
\newblock {\em International Journal of Biomathematics}, 11(03):1850046.

\bibitem[Enright and Kao, 2018]{enright2018epidemics}
Enright, J. and Kao, R.~R. (2018).
\newblock Epidemics on dynamic networks.
\newblock {\em Epidemics}, 24:88--97.

\bibitem[Grivas et~al., 2020]{grivas2020not}
Grivas, A., Alex, B., Grover, C., Tobin, R., and Whiteley, W. (2020).
\newblock Not a cute stroke: analysis of rule-and neural network-based
  information extraction systems for brain radiology reports.
\newblock In {\em Proceedings of the 11th international workshop on health text
  mining and information analysis}, pages 24--37.

\bibitem[Jim{\'e}nez-Zafra et~al., 2020]{jimenez2020corpora}
Jim{\'e}nez-Zafra, S.~M., Morante, R., Mart{\'\i}n-Valdivia, M.~T., and Lopez,
  L. A.~U. (2020).
\newblock Corpora annotated with negation: An overview.
\newblock {\em Computational Linguistics}, 46(1):1--52.

\bibitem[Krieger, 2005]{krieger2005embodiment}
Krieger, N. (2005).
\newblock Embodiment: a conceptual glossary for epidemiology.
\newblock {\em Journal of Epidemiology \& Community Health}, 59(5):350--355.

\bibitem[Levinas, 1999]{levinas1999alterity}
Levinas, E. (1999).
\newblock {\em Alterity and transcendence}.
\newblock Columbia University Press.

\bibitem[Malmqvist et~al., 2020]{malmqvist2020determining}
Malmqvist, L., Yuan, T., Nigthingale, P., and Manandhar, S. (2020).
\newblock Determining the acceptability of abstract arguments with graph
  convolutional networks.
\newblock In {\em Proceedings of the 3rd International Workshop on Systems and
  Algorithms for Formal Argumentation (SAFA’20)}.

\bibitem[Maureira et~al., 2018]{maureira2018epidemiological}
Maureira, M., Tirado, F., Torrej{\'o}n, P., and Baleriola, E. (2018).
\newblock The epidemiological factor: A genealogy of the link between medicine
  and politics.
\newblock {\em International Journal of Cultural Studies}, 21(5):505--519.

\bibitem[Nauta et~al., 2007]{nauta2007risk}
Nauta, M.~J., Jacobs-Reitsma, W.~F., and Havelaar, A.~H. (2007).
\newblock A risk assessment model for campylobacter in broiler meat.
\newblock {\em Risk Analysis: An International Journal}, 27(4):845--861.

\bibitem[Needham and Hodler, 2019]{needham2019graph}
Needham, M. and Hodler, A.~E. (2019).
\newblock {\em Graph algorithms: practical examples in Apache Spark and Neo4j}.
\newblock O'Reilly Media.

\bibitem[Negri, 2013]{negri2013spinoza}
Negri, A. (2013).
\newblock {\em Spinoza for our Time}.
\newblock Columbia University Press.

\bibitem[Nguyen-Viet et~al., 2009]{nguyen2009improving}
Nguyen-Viet, H., Zinsstag, J., Schertenleib, R., Zurbr{\"u}gg, C., Obrist, B.,
  Montangero, A., Surkinkul, N., Kon{\'e}, D., Morel, A., Ciss{\'e}, G., et~al.
  (2009).
\newblock Improving environmental sanitation, health, and well-being: a
  conceptual framework for integral interventions.
\newblock {\em EcoHealth}, 6(2):180--191.

\bibitem[O’Hare et~al., 2016]{o2016broadwick}
O’Hare, A., Lycett, S.~J., Doherty, T., M~Salvador, L.~C., and Kao, R.~R.
  (2016).
\newblock Broadwick: a framework for computational epidemiology.
\newblock {\em BMC bioinformatics}, 17(1):1--5.

\bibitem[Pagin, 2014]{pagin2014critical}
Pagin, P. (2014).
\newblock Critical pragmatics.

\bibitem[Priest, 1999]{priest1999negation}
Priest, G. (1999).
\newblock Negation as cancellation, and connexive logic.
\newblock {\em Topoi}, 18(2):141--148.

\bibitem[Priest, 2007]{priest2007paraconsistency}
Priest, G. (2007).
\newblock Paraconsistency and dialetheism.
\newblock {\em The Many Valued and Nonmonotonic Turn in Logic}, 8:129--204.

\bibitem[Racloz et~al., 2015]{racloz20158}
Racloz, V., Waltner-Toews, D., and DC, K.~S. (2015).
\newblock 8 integrated risk assessment--foodborne diseases.
\newblock {\em One Health: The Theory and Practice of Integrated Health
  Approaches}, page~85.

\bibitem[Radford et~al., 2018]{radford2018improving}
Radford, A., Narasimhan, K., Salimans, T., and Sutskever, I. (2018).
\newblock Improving language understanding by generative pre-training.

\bibitem[Raji et~al., 2021]{raji2021ai}
Raji, I.~D., Bender, E.~M., Paullada, A., Denton, E., and Hanna, A. (2021).
\newblock Ai and the everything in the whole wide world benchmark.
\newblock {\em arXiv preprint arXiv:2111.15366}.

\bibitem[Roche et~al., 2011]{roche2011agent}
Roche, B., Drake, J.~M., and Rohani, P. (2011).
\newblock An agent-based model to study the epidemiological and evolutionary
  dynamics of influenza viruses.
\newblock {\em BMC bioinformatics}, 12(1):1--10.

\bibitem[Stedden, 2020]{stedden2022}
Stedden, W. (2020).
\newblock Combining gpt-2 and bert to make a fake person.

\bibitem[Steele et~al., 2021]{STEELE2021100314}
Steele, S.~G., Toribio, J.-A.~L., and Mor, S.~M. (2021).
\newblock Global health security must embrace a one health approach:
  Contributions and experiences of veterinarians during the covid-19 response
  in australia.
\newblock {\em One Health}, 13:100314.

\bibitem[Streichert et~al., 2022]{streichert2022participation}
Streichert, L.~C., Sepe, L.~P., Jokelainen, P., Stroud, C.~M., Berezowski, J.,
  and Del Rio~Vilas, V.~J. (2022).
\newblock Participation in one health networks and involvement in the covid-19
  pandemic response: A global study.
\newblock {\em Frontiers in public health}, page 126.

\bibitem[WHO-FAO-OIE, 2019]{world2019taking}
WHO-FAO-OIE (2019).
\newblock {\em Taking a multisectoral one health approach: a tripartite guide
  to addressing zoonotic diseases in countries}.
\newblock Food \& Agriculture Org.

\bibitem[Woods et~al., 2017]{woods2017animals}
Woods, A., Bresalier, M., Cassidy, A., and Mason~Dentinger, R. (2017).
\newblock {\em Animals and the shaping of modern medicine: one health and its
  histories}.
\newblock Springer Nature.

\bibitem[Zinsstag et~al., 2020]{zinsstag2020one}
Zinsstag, J., Schelling, E., Crump, L., Whittaker, M., Tanner, M., and Stephen,
  C. (2020).
\newblock {\em One Health: the theory and practice of integrated health
  approaches}.
\newblock CABI.

\end{thebibliography}
\end{document}